# Generalized Four-momentum for Continuously Distributed Materials

**Sergey G. Fedosin**
PO box 614088, Sviazeva str. 22-79, Perm, Perm Krai, Russia
E-mail: sergey.fedosin@gmail.com

**Abstract:** A four-dimensional differential Euler-Lagrange equation for continuously distributed materials is derived based on the principle of least action, and instead of Lagrangian, this equation contains the Lagrangian density. This makes it possible to determine the density of generalized four-momentum in covariant form as derivative of the Lagrangian density with respect to four-velocity of typical particles of a system taken with opposite sign, and then calculate the generalized four-momentum itself. It is shown that the generalized four-momentum of all typical particles of a system is an integral four-vector and therefore should be considered as a special type of four-vectors. The presented expression for generalized four-momentum exactly corresponds to the Legendre transformation connecting the Lagrangian and Hamiltonian. The obtained formulas are used to calculate generalized four-momentum of stationary and moving relativistic uniform systems for the Lagrangian with particles and vector fields, including electromagnetic and gravitational fields, acceleration field and pressure field. It turns out that the generalized four-momentum of a moving system depends on the total mass of particles, on the Lorentz factor and on the velocity of the system's center of momentum. Besides, an additional contribution is made by the scalar potentials of the acceleration field and the pressure field at the center of system. The direction of the generalized four-momentum coincides with the direction of four-velocity of the system under consideration, while the generalized four-momentum is part of the relativistic four-momentum of the system.

## 1. Introduction

The three-dimensional generalized momentum is an important quantity of any system, in which fields are taken into account, since in this case the generalized momentum contains vector field potentials and replaces momentum of classical mechanics. From the standpoint of Lagrangian formalism, the generalized momentum of one particle is calculated as the partial derivative of Lagrangian with respect to velocity of this particle, and the generalized momentum of a system is equal to the sum of the generalized momenta of all system's particles [1].

In the flat Minkowski space and in curved spacetime, four-dimensional quantities are of primary importance, which requires introduction of concept of the generalized four-momentum. Unfortunately, the literature about this quantity is extremely limited. For example, in [2], a possible form of generalized four-momentum of a charged particle in external electromagnetic field is considered. The situation with calculation of generalized four-momentum, which should describe continuously distributed materials, is even worse, probably due to difficulties arising from the volume integration of physical quantities. Thus, the four-dimensional formalism for continuum mechanics is used in [3] in order to determine the relativistic stress-energy tensor and corresponding Euler-Lagrange equation in an ideal fluid. In this case, conservation laws are obtained approximately, in the form of an expansion in powers of ratio of particles' velocity to the speed of light.

In [4] it is assumed that covariant four-dimensional Euler-Lagrange equation, which is necessary to derive the generalized four-momentum, should have the following form:

$$\frac{d}{d\tau}\left(\frac{\partial L}{\partial u^{\mu}}\right)-\frac{\partial L}{\partial x^{\mu}}=0,$$

where $L$ is the Lagrangian; $u^{\mu}$ is the four-velocity; $x^{\mu}$ is the four-radius specifying the position of a particle; $d\tau = ds/c$; $c$ is the speed of light; $ds$ is the interval. The time $\tau$ is metrical proper time of the particle, which does not coincide with the physical proper time $t_p$. For differential of the time $t_p$ there is a formula [1]: $dt_p = \frac{1}{c}\sqrt{g_{00}}\,dx^0$, where $g_{00}$ is the time component of metric tensor, $dx^0 = cdt$, and $t$ is the coordinate time.

The above equation should have a different form for the case of continuous materials, because instead of the Lagrangian $L$ it is necessary to use the Lagrangian density $\mathcal{L}$, the volume integral of which gives $L$. Indeed, in order to be able to find the quantities $\frac{\partial L}{\partial u^{\mu}}$ and $\frac{\partial L}{\partial x^{\mu}}$, $L$ must depend on $u^{\mu}$ and $x^{\mu}$ of each particle in the entire set of particles included in the system. In fact, in a continuous material with many particles, the Lagrange function $L$ depends rather on the choice of observation point and on the four-velocity of typical particles at this point, than on the parameters of any specific particles, which cannot be determined due to their large number. By definition, typical particles completely characterize a physical system and are ideal statistically averaged particles. Therefore, in the case of continuous materials, all physical quantities presented in equations are calculated for typical particles, and this should also apply to the density of the generalized four-momentum.

However, we haven't found anywhere such a formula, in which the density of generalized four-momentum would be determined directly through some four-dimensional derivative of $\mathcal{L}$ for the system's typical particles. Therefore, one of our tasks will be to find the corresponding Euler-Lagrange equation for the Lagrangian density $\mathcal{L}$ and covariant expression for the density of generalized four-momentum, which is also valid in the curved spacetime.

Another of our tasks will be derivation of the generalized four-momentum in an explicit form, which would allow us to take into account all the fields of a system. In this case, we will consider four most frequently observed fields, such as electromagnetic and gravitational fields, acceleration field [5], and vector pressure field [6]. All these fields are represented as vector fields and components of a single general field [7], while gravitational field is described within the framework of covariant theory of gravitation (CTG) [8-9].

The approach used allows us to avoid difficulties that arise in the general theory of relativity (GTR) when describing motion [10]. In GTR, metric and gravitational field are merged together, so in any case it is necessary first to solve an equation for the metric in order to estimate the gravitation's contribution to the physical quantities that characterize the motion. In CTG, gravitational field exists independently of the metric, therefore in flat Minkowski space, the gravitation's contribution to material's momentum and to the acting force is taken into account exactly without solving an equation for the metric.

Our main attention will be paid to the study of possible form of generalized four-momentum arising from the Lagrangian formalism. Next, we will derive formulas for the generalized four-momentum, as well as for the relativistic momentum of the system's typical particles, and will show their relationship in the case of vector fields.

These formulas will be used to calculate the generalized four-momentum of a stationary and moving relativistic uniform system in the continuous materials limit. The choice of such physical system is not accidental, since four-potentials of fields necessary for calculating the generalized four-momentum have already been found for it by solving wave equations. In particular, expressions for scalar field potentials (32) and corresponding references are presented below in Section 6. The study of properties of the relativistic uniform system is important because such a physical system is successfully used to introduce the results of the field theory into continuum mechanics [11-12].

## 2. Lagrangian structure and its variation

By definition, the action function $S$ is integral of Lagrangian $L$ over the coordinate time $t$. In turn, the Lagrangian for continuously distributed materials in curved space-time is integral of the Lagrangian density $\mathcal{L}$ over moving volume:

$$L=\int_{V_s}\mathcal{L}\sqrt{-g}\,dx^1dx^2dx^3\,, \tag{1}$$

where $dx^1dx^2dx^3$ is product of differentials of the space coordinates, the quantity $g$ represents determinant of metric tensor $g^{\mu\nu}$, $V_s$ denotes volume of the system.

The Lagrangian density $\mathcal{L}$ is the sum of scalar terms, each of which has dimension of volumetric energy density and defines contribution to the Lagrangian density with the help of a certain energy function, associated with the corresponding field or with the four-current.

In view of (1), the action function can be represented as follows:

$$S=\int_{t_1}^{t_2}L\,dt=\int_{t_1}^{t_2}\left(\int_{V_s}\mathcal{L}\sqrt{-g}\,dx^1dx^2dx^3\right)dt\,. \tag{2}$$

Let us suppose that the Lagrangian density depends on coordinate time $t$, on the four-radius $x_n^\mu$ and on the four-velocity $u_n^\mu$ of each of $N$ system's particles with the current number $n$, on the four-potentials and field tensors at each point of the field, as well as on the metric tensor:

$$\mathcal{L}=\mathcal{L}\left(t,x_1^\mu,x_2^\mu,...x_N^\mu,u_1^\mu,u_2^\mu,...u_N^\mu,A_\mu,D_\mu,U_\mu,\pi_\mu,F_{\mu\nu},\Phi_{\mu\nu},u_{\mu\nu},f_{\mu\nu},g^{\mu\nu},R\left(g^{\mu\nu}\right)\right). \tag{3}$$

In (3), the quantities $A_\mu,D_\mu,U_\mu,\pi_\mu$ are four-potentials of electromagnetic and gravitational fields, acceleration field and pressure field, respectively, and the quantities $F_{\mu\nu},\Phi_{\mu\nu},u_{\mu\nu},f_{\mu\nu}$ are tensors of these fields. The expression $R\left(g^{\mu\nu}\right)$ means that the Lagrangian density also depends on the scalar curvature $R$, which is a function of metric tensor and its partial derivatives. In the general case, we can assume that the invariant mass density $\rho_0$ and the invariant charge density $\rho_{0q}$ of some particle with the current number $n$ are functions of time $t$ and of the four-radius $x_n^\mu$ of this particle. This leads to the fact that the mass four-current $J_n^\mu = \rho_0 u_n^\mu$, as well as the charge four-current $j_n^\mu = \rho_{0q} u_n^\mu$ of a given particle, become functions of time $t$, four-radius and four-velocity. In the final notation, the scalar terms in Lagrangian density (3) appear, as a rule, in the form of tensor invariants of the form $A_\mu j_n^\mu$, $F_{\mu\nu} F^{\mu\nu}$, and also $R = R_{\mu\nu} g^{\mu\nu}$, where $R_{\mu\nu}$ is the Ricci tensor, which is a function of the metric tensor and its derivatives.

Note that in (3) dependence of the Lagrangian density on the covariant derivatives of metric tensor is not included, since $\nabla_\lambda g^{\mu\nu} = 0$.

The variation of action function (2) is written as follows:

$$\delta S = \int_{t_1}^{t_2} \int_{V_s} \delta \mathcal{L} \sqrt{-g}\, dx^1 dx^2 dx^3 dt + \int_{t_1}^{t_2} \int_{V_s} \mathcal{L} \delta \sqrt{-g}\, dx^1 dx^2 dx^3 dt = 0 . \qquad (4)$$

The determinant $g$ of the metric tensor is a function of the metric tensor components. As a result, according to [13], the following relation holds true:

$$\delta \sqrt{-g} = -\frac{1}{2} \sqrt{-g}\; g_{\mu\nu} \delta g^{\mu\nu} . \qquad (5)$$

Let us write the variation of the Lagrangian density (3) taking into account the standard equality to zero of variation of coordinate time $\delta t = 0$:

$$\delta \mathcal{L} = \sum_{n=1}^{N} \left( \frac{\partial \mathcal{L}}{\partial x_n^\mu} \delta x_n^\mu + \frac{\partial \mathcal{L}}{\partial u_n^\mu} \delta u_n^\mu \right) + \left( \frac{\partial \mathcal{L}}{\partial A_\mu} \delta A_\mu + \frac{\partial \mathcal{L}}{\partial F_{\mu\nu}} \delta F_{\mu\nu} \right) + \left( \frac{\partial \mathcal{L}}{\partial D_\mu} \delta D_\mu + \frac{\partial \mathcal{L}}{\partial \Phi_{\mu\nu}} \delta \Phi_{\mu\nu} \right) +$$
$$+ \left( \frac{\partial \mathcal{L}}{\partial U_\mu} \delta U_\mu + \frac{\partial \mathcal{L}}{\partial u_{\mu\nu}} \delta u_{\mu\nu} \right) + \left( \frac{\partial \mathcal{L}}{\partial \pi_\mu} \delta \pi_\mu + \frac{\partial \mathcal{L}}{\partial f_{\mu\nu}} \delta f_{\mu\nu} \right) + \frac{\partial \mathcal{L}}{\partial g^{\mu\nu}} \delta g^{\mu\nu} .$$

(6)

In (6) we wrote the variation $\delta \mathcal{L}$ not in terms of covariant derivatives, but in terms of partial derivatives, using the fact that the Lagrangian density $\mathcal{L}$ is a scalar invariant, and not a four-tensor or a four-vector.

The terms in each parenthesis of (6) consists of variation of associated quantities. For example, the variation $\delta x_n^\mu$ of the four-radius of a particle with the number $n$ is related to the variation $\delta u_n^\mu$ of the four-velocity of this particle, and the variation $\delta A_\mu$ of the four-potential of electromagnetic field at an arbitrary point of the system is related to the variation $\delta F_{\mu\nu}$ of the electromagnetic field tensor. Let us substitute (5) and (6) into (4):

$$\delta S = \int_{t_1}^{t_2} \int_{V_s} \sum_{n=1}^{N} \left( \frac{\partial \mathcal{L}}{\partial x_n^\mu} \delta x_n^\mu + \frac{\partial \mathcal{L}}{\partial u_n^\mu} \delta u_n^\mu \right) \sqrt{-g}\, dx^1 dx^2 dx^3 dt +$$
$$+ \int_{t_1}^{t_2} \int_{V_s} \left[ \begin{aligned} & \left( \frac{\partial \mathcal{L}}{\partial A_\mu} \delta A_\mu + \frac{\partial \mathcal{L}}{\partial F_{\mu\nu}} \delta F_{\mu\nu} \right) + \left( \frac{\partial \mathcal{L}}{\partial D_\mu} \delta D_\mu + \frac{\partial \mathcal{L}}{\partial \Phi_{\mu\nu}} \delta \Phi_{\mu\nu} \right) + \\ & + \left( \frac{\partial \mathcal{L}}{\partial U_\mu} \delta U_\mu + \frac{\partial \mathcal{L}}{\partial u_{\mu\nu}} \delta u_{\mu\nu} \right) + \left( \frac{\partial \mathcal{L}}{\partial \pi_\mu} \delta \pi_\mu + \frac{\partial \mathcal{L}}{\partial f_{\mu\nu}} \delta f_{\mu\nu} \right) \end{aligned} \right] \sqrt{-g}\, dx^1 dx^2 dx^3 dt +$$
$$+ \int_{t_1}^{t_2} \int_{V_s} \left( \frac{\partial \mathcal{L}}{\partial g^{\mu\nu}} - \frac{1}{2} \mathcal{L} g_{\mu\nu} \right) \delta g^{\mu\nu} \sqrt{-g}\, dx^1 dx^2 dx^3 dt = 0. \tag{7}$$

Since the variables of Lagrangian (3) are independent, each integral part of (7) must vanish. In particular, the last integral vanishes under the following condition:

$$\frac{\partial \mathcal{L}}{\partial g^{\mu\nu}} - \frac{1}{2} \mathcal{L} g_{\mu\nu} = \frac{\partial \left( \mathcal{L} \sqrt{-g} \right)}{\partial g^{\mu\nu}} = 0. \tag{8}$$

The quantity $\frac{\partial \mathcal{L}}{\partial g^{\mu\nu}}$ is a derivative of the Lagrangian density with respect to the metric tensor, and Equation (8) is equation for determining the metric tensor both inside and outside the material.

In the general case, Lagrangian density (3) could also depend on the first- and even second-order partial derivatives of metric tensor with respect to coordinates and time, and these derivatives must be present in (6), (7) and (8). As a rule, all these derivatives are found only in one term of the Lagrangian density for the curved spacetime, namely in the term $\mathcal{L}_R = ckR = ckR_{\mu\nu} g^{\mu\nu}$, where $c$ is the speed of light, $k$ is a constant, $R$ is the scalar curvature, $R_{\mu\nu}$ is the Ricci tensor. However, the scalar curvature has such a property that the variation of the action function, associated with the curvature, is equal to [8], [14]:

$$\delta S_R = \int_{t_1}^{t_2} \int_{V_s} \delta \left( \mathcal{L}_R \sqrt{-g} \right) dx^1 dx^2 dx^3 dt = c\,k \int_{t_1}^{t_2} \int_{V_s} \delta \left( R_{\mu\nu}\, g^{\mu\nu} \sqrt{-g} \right) dx^1 dx^2 dx^3 dt =$$
$$= c\,k \int_{t_1}^{t_2} \int_{V_s} R_{\mu\nu}\, \delta \left( g^{\mu\nu} \sqrt{-g} \right) dx^1 dx^2 dx^3 dt.$$

From this relation we can see that the Ricci tensor $R_{\mu\nu}$, during variation of the action function with respect to the metric tensor and its first- and second-order derivatives, behaves as if it is equal to a constant, and the variation $\delta S_R$ depends only on the variation $\delta \left( g^{\mu\nu} \sqrt{-g} \right)$ with respect to the metric tensor. This justifies the form of (6), (7) and (8), and then it turns out that $\frac{\partial \mathcal{L}_R}{\partial g^{\mu\nu}} = c\,k R_{\mu\nu}$, while $\mathcal{L}_R$ is one of the terms, which are part of the Lagrangian density $\mathcal{L}$ in (8).

For the electromagnetic field, the next condition follows from (7):

$$\int_{t_1}^{t_2} \int_{V_s} \left( \frac{\partial \mathcal{L}}{\partial A_\mu} \delta A_\mu + \frac{\partial \mathcal{L}}{\partial F_{\mu\nu}} \delta F_{\mu\nu} \right) \sqrt{-g}\, dx^1 dx^2 dx^3 dt = 0 .$$

The variation $\delta F_{\mu\nu}$ should be expressed in terms of the variation $\delta A_\mu$, and after some transformations $\delta A_\mu$ should be taken outside the parentheses. What remains inside the parentheses must be equated to zero. This leads to the standard equation of electromagnetic field in the curved spacetime, which allows us to calculate the field tensor components both inside and outside the materials. Similarly, from (7) we obtain field equations for the remaining three fields.

Since in the Lagrangian density (3) the metric tensor $g^{\mu\nu}$ should not directly depend on the four-radii $x_n^\mu$ of particles, it should be assumed that $g^{\mu\nu}$ depends on $x_n^\mu$ indirectly, through other physical variables, for example, through $x^\mu$ of observation point. The difference between $x_n^\mu$ and $x^\mu$ here is that the metric tensor can be calculated at such points $x^\mu$ where there are no particles and therefore $x_n^\mu$ is not applicable.

According to an assumption in [13], in this case the four-potentials, as well as the products $j^{\mu}\sqrt{-g}$ and $J^{\mu}\sqrt{-g}$, where $j^{\mu}$ and $J^{\mu}$ denote charge and mass four-currents, respectively, do not depend directly on the metric tensor. Then, in the Equation (8) for the metric, the terms containing the products of four-potentials by four-currents and usually included in the expression for $\mathcal{L}$, disappear.

If in the Lagrangian density (3) the metric tensor directly depends on the four-radii $x_{n}^{\mu}$ of particles, then the variation $\delta g^{\mu\nu}$ in (7) must be expressed through the variations $\delta x_{n}^{\mu}$ of the particles. Then one should transform the last integral in (7), single out $\delta x_{n}^{\mu}$ separately and connect this integral to the first integral in (7) to find the equation of motion.

Another equivalent approach assumes [5] that Lagrangian density (3) instead of four-radii $x_{n}^{\mu}$ and four-velocities $u_{n}^{\mu}$ of individual particles directly depends on four-currents $j^{\mu}$ and $J^{\mu}$ in the following form:

$$\mathcal{L}=\mathcal{L}\left[t, j^{\mu}\left(x^{\mu},u^{\mu}\right), J^{\mu}\left(x^{\mu},u^{\mu}\right), A_{\mu}, D_{\mu}, U_{\mu}, \pi_{\mu}, F_{\mu\nu}, \Phi_{\mu\nu}, u_{\mu\nu}, f_{\mu\nu}, g^{\mu\nu}, R\left(g^{\mu\nu}\right)\right], \qquad (9)$$

where $x^{\mu}$ specifies the observation point, $u^{\mu}$ is the four-velocity of a typical particle at that point.

This leads to the fact that instead of equation of motion with a generalized momentum, the equation of motion of particles with field tensors and four-currents appears, while the Equation (8) for the metric and the equations for determining the field tensors remain valid. From this it follows that the Equation (8) for metric should be fulfilled regardless of how the metric tensor depends on physical variables, including dependencies on $x_{n}^{\mu}$ and $u_{n}^{\mu}$ of individual particles, or dependence on $x^{\mu}$ of observation point.

## 3. Four-dimensional Euler-Lagrange equation

We can assume that all the arguments in the previous section refer to typical particles, the set of which continuously fills a certain volume and represents the material of a physical system. In this case, the difference between observation point given by the four-vector $x^{\mu}$ and the four-radius $x_{n}^{\mu}$ of a typical particle at this point disappears, and the four-velocity $u^{\mu}$ of a typical particle at observation point is equal to $u_{n}^{\mu}$. According to expression (9) of the Lagrangian density, only the charge and mass four-currents $j^{\mu}$ and $J^{\mu}$, which are present in each Lagrangian density for continuous materials, can be direct functions of the observation point $x^\mu$ and the four-velocity $u^\mu$ of a typical particle of material at this point. As for the four-potentials and the field tensors, as well as the metric tensor, they become functions of $x^\mu$ and $u^\mu$ only after the corresponding field equations are solved. Let us denote the sum of Lagrangian density terms, containing the four-currents, by $\mathcal{L}_p$. The first integral in (7) must be equal to zero irrespectively of the other integrals, and we should substitute $\mathcal{L}_p$ into it as a part of Lagrangian density, containing dependence on $x^\mu$ and $u^\mu$ of typical particles:

$$\delta S_p = \int_{t_1}^{t_2}\int_{V_s}\sum_{n=1}^{N}\left(\frac{\partial \mathcal{L}_p}{\partial x_n^\mu}\delta x_n^\mu + \frac{\partial \mathcal{L}_p}{\partial u_n^\mu}\delta u_n^\mu\right)\sqrt{-g}\, dx^1 dx^2 dx^3 dt \approx$$
$$\approx \frac{1}{c}\int_{t_1}^{t_2}\int_{V_s}\sum_{n=1}^{N}\left(\frac{\partial \mathcal{L}_p}{\partial x_n^\mu}\delta x_n^\mu + \frac{\partial \mathcal{L}_p}{\partial u_n^\mu}\delta u_n^\mu\right)d\Omega_n = \sum_{n=1}^{N}\int_{\tau_1}^{\tau_2}\int_{V_n}\left(\frac{\partial \mathcal{L}_p}{\partial x_n^\mu}\delta x_n^\mu + \frac{\partial \mathcal{L}_p}{\partial u_n^\mu}\delta u_n^\mu\right)dV_n\, d\tau_n = 0, \tag{10}$$

where the element of covariant four-volume $\sqrt{-g}\, dx^0 dx^1 dx^2 dx^3 = c dV_n d\tau_n = d\Omega_n$ with the current number $n$ is present, $dx^0 = c\,dt$, $c$ is the speed of light, $\tau_n$ is the proper time of typical particle with the number $n$, $dV_n$ is the particle's proper volume.

In (10) the sum must be integrated over the entire volume $V_s$ of the system, while each term of the sum is associated with only one particle. This means that in (10) we can go from the integral over the entire volume $V_s$ to the sum of the integrals over the volumes of individual typical particles, while leaving the integrand unchanged. We reflected this with the help of last two terms in (10).

The proper time $\tau_n$ of any particle in (10) is not equal to the proper time of any other particle in the system. Based on this, it is believed that it is possible to derive the Euler-Lagrange equation in a covariant form only for one particle, but for a system of particles it is impossible. Probably, this explains the absence in literature of a covariantly defined four-vector of the generalized momentum density for a continuous material. Thus, we come to conclusion that it is necessary to change the procedure of variation and adapt it to the case under consideration. Let's do it as follows.

The variations $\delta x^\mu$ can be considered as small acceptable deviations from the true trajectory of a particle under consideration, moving in space and time between two given points. We will take into account definition of the four-velocity $u^{\mu}=\frac{dx^{\mu}}{d\tau}$, and will define its variation as follows: $\delta u^{\mu}=\frac{\delta(dx^{\mu})}{d\tau}=\frac{d(\delta x^{\mu})}{d\tau}$. Despite the fact that the proper time $\tau$ in each particle flows at different speeds, further we will assume synchronization of variation with respect to proper time for all particles. To do this, it suffices to synchronize the origin of the particles' proper time and perform variation at this moment. The entire time interval $t_2-t_1$, within which the time integration is performed in (10), corresponds to a certain time interval $\tau_2-\tau_1$ for the particle with the current number $n$, and the interval $\tau_2-\tau_1$ will be different for different particles. The interval $t_2-t_1$ during integration in (10) is divided into a set of time differentials $dt$, similarly, for each particle the corresponding interval $\tau_2-\tau_1$ is divided into a set of time differentials $d\tau$. Since the four-velocity $u^{\mu}$ of each particle is constantly changing, within each differential $d\tau$ at the time point $\tau$ the particle would have a different four-velocity $u^{\mu}$ and a different time component of the four-velocity $u^0$. Thus, in order to sufficiently accurately cover all the trajectories of the system's particles during the action variation with respect to the proper time, it is necessary to synchronize the origin of the proper time $\tau$ of all the particles many times, within each of the corresponding time differentials $d\tau$.

On the other hand, in view of the relation $u^0=c\frac{dt}{d\tau}$ we can write the following:

$$\delta u^{\mu}=\delta\left(\frac{u^0}{c}\frac{dx^{\mu}}{dt}\right)=\frac{dx^{\mu}}{dt}\delta\left(\frac{u^0}{c}\right)+\frac{u^0}{c}\frac{d(\delta x^{\mu})}{dt}.$$

If we assume $\delta u^0=0$ here, then variation of the four-velocity will reduce to the value $\delta u^{\mu}=\frac{d(\delta x^{\mu})}{d\tau}$ provided above. Thus, we will assume that within each time differential with the duration $dt$ neither the time $t$, nor the time component $u^0$ of four-velocity is varied, behaving as a constant value within this differential. In this case, the component $u^0$ within different differentials $dt$, that is, at different time points, may differ in value, changing its value in a stepwise fashion during transition to a new time differential.

With this in mind, we will transform the last expression in (10) by parts for each particle:

$$\delta S_p \approx \sum_{n=1}^{N}\int_{\tau_1}^{\tau_2}\int_{V_n}\left(\frac{\partial \mathcal{L}_p}{\partial x_n^\mu}\delta x_n^\mu + \frac{\partial \mathcal{L}_p}{\partial u_n^\mu}\delta u_n^\mu\right)dV_n\,d\tau_n = \sum_{n=1}^{N}\int_{\tau_1}^{\tau_2}\int_{V_n}\left(\frac{\partial \mathcal{L}_p}{\partial x_n^\mu}\delta x_n^\mu + \frac{\partial \mathcal{L}_p}{\partial u_n^\mu}\frac{d(\delta x_n^\mu)}{d\tau_n}\right)dV_n\,d\tau_n =$$
$$= \sum_{n=1}^{N}\int_{\tau_1}^{\tau_2}\int_{V_n}\left[\frac{\partial \mathcal{L}_p}{\partial x_n^\mu} - \frac{d}{d\tau_n}\left(\frac{\partial \mathcal{L}_p}{\partial u_n^\mu}\right)\right]\delta x_n^\mu\, dV_n\,d\tau_n + \sum_{n=1}^{N}\int_{\tau_1}^{\tau_2}\int_{V_n}\frac{d}{d\tau_n}\left(\frac{\partial \mathcal{L}_p}{\partial u_n^\mu}\delta x_n^\mu\right)dV_n\,d\tau_n = 0. \qquad (11)$$

Let us now transform the last term in (11):

$$\sum_{n=1}^{N}\int_{\tau_1}^{\tau_2}\int_{V_n}\frac{d}{d\tau_n}\left(\frac{\partial \mathcal{L}_p}{\partial u_n^\mu}\delta x_n^\mu\right)dV_n\,d\tau_n = \sum_{n=1}^{N}\int_{V_n}\left(\frac{\partial \mathcal{L}_p}{\partial u_n^\mu}\delta x_n^\mu\right)\Bigg|_{\tau_1}^{\tau_2} dV_n = 0\,.$$

In this equation inside the volume integral of particle with the number $n$, the variations $\delta x_n^\mu(\tau_1)$ at the initial time points $t_1$ and $\tau_1$, and the variations $\delta x_n^\mu(\tau_2)$ at the final time points $t_2$ and $\tau_2$ are equal to zero by the condition of variation. As a result, the last term in (11) vanishes and the following remains:

$$\delta S_p \approx \sum_{n=1}^{N}\int_{\tau_1}^{\tau_2}\int_{V_n}\left[\frac{\partial \mathcal{L}_p}{\partial x_n^\mu} - \frac{d}{d\tau_n}\left(\frac{\partial \mathcal{L}_p}{\partial u_n^\mu}\right)\right]\delta x_n^\mu\, dV_n\,d\tau_n = 0\,.$$

In the general case, the variations $\delta x_n^\mu$ are different for different particles, do not depend on each other, are arbitrary and non-zero. In order for the above relation to hold, the expression in the square brackets under the summation sign must be equal to zero. Hence, we obtain the four-dimensional Euler-Lagrange equation for each of the particles:

$$\frac{\partial \mathcal{L}_p}{\partial x_n^\mu} - \frac{d}{d\tau_n}\left(\frac{\partial \mathcal{L}_p}{\partial u_n^\mu}\right) = 0\,. \qquad (12)$$

On the other hand, within volume of one particle and during the time differential $d\tau_n$, the time component $u_n^0 = c\frac{dt}{d\tau_n}$ of four-velocity of the particle remains constant, according to the condition of variation that we have accepted, and it can be introduced under the derivative sign $\frac{\partial \mathcal{L}_p}{\partial x_n^\mu}$. Let us multiply and at the same time divide by $u_n^0$ the expression inside the integral for $\delta S_p$:

$$\delta S_p \approx \sum_{n=1}^{N} \int_{\tau_1}^{\tau_2} \int_{V_n} \left[ \frac{\partial}{\partial x_n^\mu} \left( \frac{\mathcal{L}_p}{u_n^0} \right) - \frac{1}{u_n^0} \frac{d}{d\tau_n} \left( \frac{\partial \mathcal{L}_p}{\partial u_n^\mu} \right) \right] u_n^0 \delta x_n^\mu \, dV_n \, d\tau_n =$$
$$= \sum_{n=1}^{N} \int_{\tau_1}^{\tau_2} \int_{V_n} \left[ \frac{\partial}{\partial x_n^\mu} \left( \frac{\mathcal{L}_p}{u_n^0} \right) - \frac{1}{c} \frac{d}{dt} \left( \frac{\partial \mathcal{L}_p}{\partial u_n^\mu} \right) \right] u_n^0 \delta x_n^\mu \, dV_n \, d\tau_n = 0.$$

Since in this expression the quantities $u_n^0$ and $\delta x_n^\mu$ in the general case are arbitrary and non-zero, the following relation must hold true in the first approximation:

$$\frac{\partial}{\partial x_n^\mu} \left( \frac{\mathcal{L}_p}{u_n^0} \right) - \frac{1}{c} \frac{d}{dt} \left( \frac{\partial \mathcal{L}_p}{\partial u_n^\mu} \right) = 0 .$$

Let us denote volumetric density of the generalized four-momentum by $\mathcal{P}_\mu = -\frac{\partial \mathcal{L}_p}{\partial u^\mu}$. Taking this into account, removing the particle's number $n$, we arrive at a relation, which corresponds by its form to the differential equation of motion of a typical particle:

$$\frac{d\mathcal{P}_\mu}{dt} = -c \frac{\partial}{\partial x^\mu} \left( \frac{\mathcal{L}_p}{u^0} \right) = \mathcal{F}_\mu . \qquad (13)$$

The Euler-Lagrange Equation (12) was obtained under the condition that the time component $u_n^0 = c \frac{dt}{d\tau_n}$ is constant, which corresponds to motion of particles at a constant speed, the value of which depends on selected time differential $dt$ within the time interval $t_2 - t_1$. In the limit of continuously distributed materials, particles cannot move in such way due to continuous interactions with each other, therefore instead of (12) we will use (13) as the most appropriate expression in this case.

A feature of Equations (12) and (13) is that they are expressed in terms of derivatives of the Lagrangian density $\mathcal{L}$, and not in terms of derivatives of the Lagrangian $L$. It should be noted that Equations (12) and (13) are valid to the same extent, since they used the same condition for the constancy of time component $u_n^0 = c\frac{dt}{d\tau_n}$ of typical particles upon variation of Lagrangian density. The less the differences $t_2 - t_1$ and $\tau_2 - \tau_1$ are in (10), the better the condition $u_n^0 = const$ is satisfied during variation, and the more precisely we can state that Equations (12) and (13) are valid, including in curved spacetime.

The structure of Equations (12) and (13) is such that they represent one of the possible forms of four-dimensional equations of motion of typical particles. In this case, on the left side of (13) there is a full rate of change with time of the density of generalized four-momentum, respectively, on the right side there is the volume density of generalized four-force $\mathcal{F}_\mu$.

The equation of motion can be written in at least three more equivalent forms, for example, in terms of field tensors, in terms of field four-potentials, and in terms of energy-momentum tensors of fields [5]. Thus, in [11] a covariant equation of motion, valid in a curved space-time, was derived from the principle of least action, taking into account dissipation vector field, pressure field, acceleration field, gravitational and electromagnetic fields. This equation, expressed in terms of field tensors and four-currents, accurately reproduces the Navier-Stokes equation in the limit of weak field.

### 4. Generalized four-momentum

Let us suppose now that all the system's particles are simultaneously shifted by a certain constant four-vector $\delta x^\mu = \varepsilon^\mu = const$, which is a variation of the four-radius $x^\mu$. Since

$$\delta u^\mu = \delta\left(\frac{u^0}{c}\frac{dx^\mu}{dt}\right) = \frac{dx^\mu}{dt}\delta\left(\frac{u^0}{c}\right) + \frac{u^0}{c}\delta\left(\frac{dx^\mu}{dt}\right) = \frac{dx^\mu}{c}\frac{\delta u^0}{dt} + \frac{u^0}{c}\frac{d\varepsilon^\mu}{dt},$$

then if $u^0 = const$, $\delta u^0 = 0$, $\varepsilon^\mu = const$, $d\varepsilon^\mu = 0$, here $\delta u^\mu = 0$ will be. In this case, the variation $\delta x^\mu$ leads in (10) to the variation of action function of the following form:

$$\begin{aligned}\delta S_1 &\approx \sum_{n=1}^{N}\int_{\tau_1}^{\tau_2}\int_{V_n}\frac{\partial \mathcal{L}_p}{\partial x_n^\mu}\delta x^\mu\, dV_n\, d\tau_n = \varepsilon^\mu\sum_{n=1}^{N}\int_{\tau_1}^{\tau_2}\int_{V_n}\frac{d}{d\tau_n}\left(\frac{\partial \mathcal{L}_p}{\partial u_n^\mu}\right)dV_n\, d\tau_n = \\ &= \frac{1}{c}\varepsilon^\mu\sum_{n=1}^{N}\int_{\tau_1}^{\tau_2}\int_{V_n}u_n^0\frac{d}{dt}\left(\frac{\partial \mathcal{L}_p}{\partial u_n^\mu}\right)dV_n\, d\tau_n \approx \frac{1}{c}\varepsilon^\mu\sum_{n=1}^{N}\int_{\tau_1}^{\tau_2}\frac{d}{dt}\int_{V_n}\left(\frac{\partial \mathcal{L}_p}{\partial u_n^\mu}u_n^0\right)dV_n\, d\tau_n = 0.\end{aligned} \tag{14}$$

In order to transform the sum in (14), we used (12) and the expression $\frac{dt}{d\tau_n}=\frac{1}{c}u_n^0$. It is also assumed that when integrating over proper time and over volume of one particle, the value $u_n^0$ is constant and, on the average, does not depend on the time, just as in an equilibrium system, so that $u_n^0$ can be introduced under the time derivative sign. In the limit of continuous materials we can go over from the sum of integrals over the volume of individual particles to one integral over the entire system's volume, for which in the right-hand side of (14) we can replace the product of differentials $dV_n d\tau_n$ by $\sqrt{-g}\,dx^1dx^2dx^3dt$, similarly to (10), and remove the particles' number $n$:

$$\delta S_1 \approx \frac{1}{c}\varepsilon^{\mu}\int_{t_1}^{t_2}\frac{d}{dt}\left\{\int_{V_s}\frac{\partial \mathcal{L}_p}{\partial u^{\mu}}u^0\sqrt{-g}\,dx^1dx^2dx^3\right\}dt = 0\,. \qquad (15)$$

Let us now introduce for consideration the generalized four-momentum of the system's particles:

$$p_{\mu} = -\frac{1}{c}\int_{V_s}\frac{\partial \mathcal{L}_p}{\partial u^{\mu}}u^0\sqrt{-g}\,dx^1dx^2dx^3 = \frac{1}{c}\int_{V_s}\mathcal{P}_{\mu}u^0\sqrt{-g}\,dx^1dx^2dx^3 = \int_{V_{s0}}\mathcal{P}_{\mu}\,dV_0\,. \qquad (16)$$

In (16), we used definition of the density of generalized four-momentum $\mathcal{P}_{\mu}=-\frac{\partial \mathcal{L}_p}{\partial u^{\mu}}$ from (13) and the relation from [1]:

$$\frac{dt}{d\tau}\sqrt{-g}\,dx^1dx^2dx^3 = \frac{u^0}{c}\sqrt{-g}\,dx^1dx^2dx^3 = dV_0\,, \qquad (17)$$

where $dV_0$ is the differential of the proper volume of any of particles, calculated in the particle's comoving reference frame.

We should note that by its construction method $p_{\mu}=\int_{V_{s0}}\mathcal{P}_{\mu}\,dV_0$ is a four-vector, as well as $\mathcal{P}_{\mu}$. Besides, it is assumed that at the time of calculation of both four-vectors, the time components of four-velocities of all the particles either do not change or are averaged over time.

The fact that the density of generalized four-momentum is a four-vector is obvious from the covariant definition $\mathcal{P}_\mu = -\frac{\partial \mathcal{L}_p}{\partial u^\mu}$. In the limit of continuously distributed materials, we can assume that typical particles almost completely fill entire volume of the system. Then the generalized four-momentum $p_\mu$ in (16) is obtained equal to the integral sum of the products of $\mathcal{P}_\mu$ of individual particles by the invariant volumes of these particles. Since the product of a scalar by a four-vector gives a four-vector, and the sum of four-vectors is a four-vector, the generalized four-momentum $p_\mu$ in (16) is a four-vector.

A relation for the generalized four-momentum follows from (15) and (16):

$$\frac{dp_\mu}{dt} = 0. \qquad (18)$$

According to (18), if shifting of all the system's particles by the constant four-vector $\delta x^\mu = \varepsilon^\mu$ does not change physical properties of the system, then hence it follows that the generalized four-momentum $p_\mu$ is conserved. A closed system does not depend on environment and on fields from external sources, and for it the condition of the system's constancy during the particles' transfer is satisfied. Therefore, for a closed system the relation $p_\mu = const$ will be valid.

It should be noted that this transfer by the constant four-vector $\delta x^\mu = \varepsilon^\mu$ should be considered as part of process of variation of variables, and not as a real process of the particles' motion, in which the periods of acceleration and emission of charged particles are inevitable, which leads to a change in balance of energy and momentum, changes physical properties of the system, and violates conditions of variation.

In (14), we assumed that $u_n^0$ is a constant value when integrated over volume of each typical particle and, on the average, does not depend on time. But this is precisely what is characteristic of an equilibrium system described with the help of typical particles and the procedure of averaging physical quantities, and this fully justifies our approach.

In this case, we can go further and introduce $u_n^0$ under the partial derivative sign in (14), taking into account $\delta x^\mu = \varepsilon^\mu$ and then replacing the product of differentials $dV_n d\tau_n$ by $\sqrt{-g}\,dx^1 dx^2 dx^3 dt$, similarly to (10), and going over to the approximation of continuous materials:

$$\delta S_1 \approx \frac{1}{c}\delta x^\mu \sum_{n=1}^{N}\int_{\tau_1}^{\tau_2}\int_{V_n} u_n^0 \frac{d}{dt}\left(\frac{\partial \mathcal{L}_p}{\partial u_n^\mu}\right) dV_n\, d\tau_n \approx \frac{1}{c}\delta x^\mu \sum_{n=1}^{N}\int_{\tau_1}^{\tau_2}\frac{d}{dt}\int_{V_n}\frac{\partial\left(\mathcal{L}_p/u_n^0\right)}{\partial\left(u_n^\mu/u_n^0\right)} u_n^0 dV_n\, d\tau_n \approx$$
$$\approx \frac{1}{c}\delta x^\mu \int_{t_1}^{t_2}\frac{d}{dt}\left\{\int_{V_s}\frac{\partial\left(\mathcal{L}_p/u^0\right)}{\partial\left(u^\mu/u^0\right)} u^0 \sqrt{-g}\, dx^1 dx^2 dx^3\right\} dt = 0. \qquad (19)$$

Let us define a new four-dimensional quantity:

$$\Im_\mu = -\frac{1}{c}\int_{V_s}\frac{\partial\left(\mathcal{L}_p/u^0\right)}{\partial\left(u^\mu/u^0\right)} u^0 \sqrt{-g}\, dx^1 dx^2 dx^3 . \qquad (20)$$

The relation $\frac{d\Im_\mu}{dt}=0$ follows from (19) and (20), that is, $\Im_\mu = const$ for an equilibrium closed system.

In the general case, $\Im_\mu$ is not a four-vector, but becomes it on condition that $u^0$ for each particle does not change at the moment of calculating $\Im_\mu$. Indeed, in this case the relation $\frac{\partial\left(\mathcal{L}_p/u^0\right)}{\partial\left(u^\mu/u^0\right)} = \frac{\partial \mathcal{L}_p}{\partial u^\mu}$ will be satisfied, and then $\Im_\mu$ becomes equal to the generalized four-momentum $p_\mu$ in (16).

The significance of $\Im_\mu$ in (20) lies in the fact that its space component up to a sign equals the relativistic momentum of system's particles. In order to see this, we will take into account the following relations: $x^\mu = (ct, \mathbf{r})$, $\frac{dt}{d\tau} = \frac{1}{c}u^0$, $u^\mu = \frac{dx^\mu}{d\tau} = \frac{1}{c}u^0 \frac{dx^\mu}{dt} = \frac{1}{c}u^0 (c, \mathbf{v})$. If we set $\Im_\mu = (\Im_0, -\boldsymbol{\Im})$, then for $\boldsymbol{\Im}$ from (20) it follows:

$$\boldsymbol{\Im} = \int_{V_s}\frac{\partial\left(\mathcal{L}_p/u^0\right)}{\partial \mathbf{v}} u^0 \sqrt{-g}\, dx^1 dx^2 dx^3 . \qquad (21)$$

In [1], the three-dimensional generalized momentum of a system, which takes into account all the acting fields and actually represents the total relativistic momentum of the system's particles, is determined as follows:

$$\mathbf{p}=\sum_{n=1}^{N}\frac{\partial L}{\partial \mathbf{v}_n}.$$

We must again take into account our reasoning in Section 3 about dependence of the Lagrangian density on time, coordinates, and the particles' velocities. Only the part of the Lagrangian density, which we have denoted by $\mathcal{L}_p$ and which contains the four-currents, can directly depend on the particles' velocities. With this in mind, and in view of relations (1) and (17), we find:

$$L_p=\int_{V_s}\mathcal{L}_p\sqrt{-g}\,dx^1dx^2dx^3=c\int_{V_{s0}}\frac{\mathcal{L}_p}{u^0}dV_0=c\sum_{n=1}^{N}\int_{V_n}\frac{\mathcal{L}_p}{u_n^0}dV_n\ ,$$

$$\mathbf{p}=\sum_{n=1}^{N}\frac{\partial L_p}{\partial \mathbf{v}_n}=c\sum_{n=1}^{N}\frac{\partial}{\partial \mathbf{v}_n}\left(\int_{V_n}\frac{\mathcal{L}_p}{u_n^0}dV_n\right)=c\sum_{n=1}^{N}\int_{V_n}\frac{\partial}{\partial \mathbf{v}_n}\left(\frac{\mathcal{L}_p}{u_n^0}\right)dV_n=\int_{V_s}\frac{\partial\left(\mathcal{L}_p/u^0\right)}{\partial \mathbf{v}}u^0\sqrt{-g}\,dx^1dx^2dx^3. \tag{22}$$

The obtained expression coincides with $\mathbf{\mathfrak{I}}$ in (21), so that $\mathbf{p}=\mathbf{\mathfrak{I}}$. If we denote the generalized four-momentum in the form $p_\mu=(p_0,-\mathbf{p})$ and take into account the coincidence $p_\mu$ and $\mathfrak{I}_\mu=(\mathfrak{I}_0,-\mathbf{\mathfrak{I}})$ provided that for each particle $u^0$ does not change at the moment the momentum is calculated, then $\mathbf{p}$ will be both the total relativistic momentum of the particles of the system in (22) and the total generalized momentum of the particles included in $p_\mu$ (16).

As for the obtaining procedure and physical meaning of the four-vectors $\mathcal{P}_\mu$ and $p_\mu$, a few remarks should be made. First of all, the displacement of all the system's particles, without exception, to a certain constant four-vector $\delta x^\mu=\varepsilon^\mu$ in one direction, which leaves the physical system unchanged and is presented in (14), is closely related to Noether's theorem. Only with such a displacement, it is guaranteed that the system would preserve its form, relative position of the particles and their velocities, as well as the fields' magnitudes, which would lead to the momentum conservation. In symmetric systems other displacements are possible, for example, inversion of the coordinates of all the particles (parity transformation) or substitution of the opposite particles with each other. According to Noether's theorem, each continuous symmetry corresponds to its own transformation of the particles' coordinates and its own conservation law of one or another physical quantity.

Secondly, in the approximation of continuous materials, in the equations, instead of the Lagrangian $L$, it is convenient to use its volumetric density $\mathcal{L}$, which allows us to refuse from integration in (1). Thirdly, due to the large number of interacting particles, the four-potentials and tensors of fields acting in the material no longer depend on coordinates and velocities of individual particles, they are determined only by the properties of the system as a whole, and in the center-of-momentum frame they depend mainly on coordinates of the observation point.

As for the charge (electromagnetic) and mass four-currents that are also part of the Lagrangian density, it is believed that these four-currents are associated with the motion of the so-called typical particles of the system. The characteristic of typical particles is that they define the basic features of the physical system and allow it to be described in the most complete way. The independence of field functions from the coordinates of individual particles and the emergence of typical particles take place during averaging of motions of individual particles and gauging of the properties of these particles. As a result of such averaging, we can assume that at a certain point in the stationary equilibrium system, typical particles move at a certain averaged four-velocity $u^{\mu}$, depending on the coordinates of observation point. The time component $u^0$ of four-velocity of typical particles can also be considered averaged, moreover, in the stationary system as a whole, $u^0$ will be constant, although it will differ in value in different parts of the system. It is this constancy of $u^0$ of typical particles that can be implied in the derivation of Equations (12) and (13), and (12) and (13) can be considered as equations for averaged physical quantities. Another way to imagine the constancy of $u^0$, necessary to derive the generalized four-momentum density $\mathcal{P}_{\mu}$ in (13), is to assume that $u^0$ is calculated as an instantaneous value per short time, during which the velocities of typical particles do not have enough time to change significantly. Thus, we can consider our approach to be valid at least for systems that are in equilibrium and consist of a continuously distributed material. In Section 8, we will also show that the generalized four-momentum concept presented by us is consistent with both Hamiltonian mechanics and Lagrangian mechanics.

The peculiar feature of the generalized four-momentum $p_{\mu}$ in (16) is unusual method of its determination in terms of volume integral. Indeed, the standard four-vectors are defined locally or in a point volume, which allows us to make transitions from the form with a covariant index to the form with a contravariant index using the metric tensor at a given point, for example, $A^{\mu} = g^{\mu\nu} A_{\nu}$, $A_{\mu} = g_{\mu\nu} A^{\nu}$. However, $p_{\mu}$ defines the generalized four-momentum for all the particles and is calculated as the integral over a sufficiently large volume. Such four-vectors are not local and should be called integral four-vectors. For such four-vectors, the equality of the type $p^{\mu}=g^{\mu\nu}p_{\nu}$ in the general case will not hold true, since the metric tensor $g^{\mu\nu}$ can have different values at each point of the system. In order to obtain the contravariant form of $p^{\mu}$, we should turn to definition of the integral four-vector in (16):

$$p^{\mu}=\int_{V_{s0}}\mathcal{P}^{\mu}dV_{0}=\frac{1}{c}\int_{V_{s}}g^{\mu\nu}\mathcal{P}_{\nu}u^{0}\sqrt{-g}\,dx^{1}dx^{2}dx^{3}\,.$$

## 5. Lagrangian density for vector fields

In order to calculate the generalized four-momentum, we will use the Lagrangian density for four vector fields in a curved space-time, according to [5], [11]:

$$\begin{aligned}\mathcal{L}=&-A_{\mu}\,j^{\mu}-D_{\mu}\,J^{\mu}-U_{\mu}\,J^{\mu}-\pi_{\mu}\,J^{\mu}-\frac{1}{4\mu_{0}}F_{\mu\nu}F^{\mu\nu}+\frac{c^{2}}{16\pi G}\Phi_{\mu\nu}\Phi^{\mu\nu}-\\&-\frac{c^{2}}{16\pi\eta}u_{\mu\nu}u^{\mu\nu}-\frac{c^{2}}{16\pi\sigma}f_{\mu\nu}f^{\mu\nu}+ckR-2ck\Lambda,\end{aligned}\qquad(23)$$

where $A_{\mu}=\left(\frac{\varphi}{c},-\mathbf{A}\right)$ is the four-potential of electromagnetic field, defined by the scalar potential $\varphi$ and the vector potential $\mathbf{A}$ of this field,

$j^{\mu}=\rho_{0q}u^{\mu}$ is the charge four-current,

$\rho_{0q}$ is the charge density in the particle's comoving reference frame,

$u^{\mu}$ is the four-velocity of a point particle,

$D_{\mu}=\left(\frac{\psi}{c},-\mathbf{D}\right)$ is the four-potential of gravitational field, described by the scalar potential $\psi$ and the vector potential $\mathbf{D}$ of this field within the framework of the covariant theory of gravitation,

$J^{\mu}=\rho_{0}u^{\mu}$ is the mass four-current,

$\rho_{0}$ is the mass density in the particle's comoving reference frame,

$U_{\mu}=\left(\frac{\vartheta}{c},-\mathbf{U}\right)$ is the four-potential of acceleration field, where $\vartheta$ and $\mathbf{U}$ denote the scalar and vector potentials, respectively, $\pi_\mu = \left( \frac{\wp}{c}, -\mathbf{\Pi} \right)$ is the four-potential of pressure field, consisting of the scalar potential $\wp$ and the vector potential $\mathbf{\Pi}$; if inside the particle the vector potential of pressure field is equal to zero, then $\pi_\mu = \frac{p_0}{\rho_0 c^2} u_\mu$, where $p_0$ is the pressure in the particle's comoving reference frame,

$\mu_0$ is the magnetic constant,

$F_{\mu\nu} = \nabla_\mu A_\nu - \nabla_\nu A_\mu = \partial_\mu A_\nu - \partial_\nu A_\mu$ is the electromagnetic tensor,

$G$ is the gravitational constant,

$\Phi_{\mu\nu} = \nabla_\mu D_\nu - \nabla_\nu D_\mu = \partial_\mu D_\nu - \partial_\nu D_\mu$ is the gravitational tensor,

$\eta$ is the acceleration field coefficient,

$u_{\mu\nu} = \nabla_\mu U_\nu - \nabla_\nu U_\mu = \partial_\mu U_\nu - \partial_\nu U_\mu$ is the acceleration tensor, calculated as the four-curl of the four-potential of acceleration field,

$\sigma$ is the pressure field coefficient,

$f_{\mu\nu} = \nabla_\mu \pi_\nu - \nabla_\nu \pi_\mu = \partial_\mu \pi_\nu - \partial_\nu \pi_\mu$ is the pressure field tensor,

$k = -\frac{c^3}{16\pi G \beta}$, where $\beta$ is some coefficient of the order of unity to be determined,

$R$ is the scalar curvature,

$\Lambda$ is the cosmological constant.

The charge density $\rho_{0q}$ and mass density $\rho_0$ included in the corresponding four-currents are not constants and they defined as covariant scalar functions of four-radii and four-momenta of typical particles of a system. This means that when the Lagrangian density (23) is varied in principle of least action, $\rho_{0q}$ and $\rho_0$ must also be varied, such as, for example, the scalar curvature $R$.

According to [5], in order to gauge the relativistic energy of a system, the cosmological constant is defined in such a way that the condition $R = 2\Lambda$ arises. In this case, the energy will not depend on $R$ and $\Lambda$ and becomes uniquely defined. The same applies to the generalized four-momentum. Therefore, when calculating it, we will assume that in (23) $R = 2\Lambda$. Then from (23) the expression follows for $\mathcal{L}_p$ as that part of the Lagrangian density, which contains four-currents as functions of the four-radius $x^\mu$ and four-velocity $u^\mu$ of an arbitrary typical particle:

$$\mathcal{L}_p = -A_\mu\, j^\mu - D_\mu\, J^\mu - U_\mu\, J^\mu - \pi_\mu\, J^\mu . \tag{24}$$

In the simplest case, when the global four-potentials and field tensors do not depend on four-velocities of individual system's particles, the density of generalized four-momentum for $\mathcal{L}_p$ (24) will be equal to:

$$\mathcal{P}_\mu = -\frac{\partial \mathcal{L}_p}{\partial u^\mu} = \rho_{0q} A_\mu + \rho_0 D_\mu + \rho_0 U_\mu + \rho_0 \pi_\mu . \tag{25}$$

From (16) the expression follows for the generalized four-momentum in this case:

$$p_\mu = \frac{1}{c}\int_{V_s} \left(\rho_{0q} A_\mu + \rho_0 D_\mu + \rho_0 U_\mu + \rho_0 \pi_\mu\right) u^0 \sqrt{-g}\, dx^1 dx^2 dx^3 . \tag{26}$$

Since $p_\mu = (p_0, -\mathbf{p})$, for the generalized momentum we find:

$$\mathbf{p} = \frac{1}{c}\int_{V_s} \left(\rho_{0q}\mathbf{A} + \rho_0 \mathbf{D} + \rho_0 \mathbf{U} + \rho_0 \mathbf{\Pi}\right) u^0 \sqrt{-g}\, dx^1 dx^2 dx^3 . \tag{27}$$

For $\mathfrak{J}$ (21), in view of (24), we obtain the following:

$$\mathfrak{J} = \int_{V_s} \frac{\partial\left(\mathcal{L}_p / u^0\right)}{\partial \mathbf{v}} u^0 \sqrt{-g}\, dx^1 dx^2 dx^3 = \frac{1}{c}\int_{V_s} \left(\rho_{0q}\mathbf{A} + \rho_0 \mathbf{D} + \rho_0 \mathbf{U} + \rho_0 \mathbf{\Pi}\right) u^0 \sqrt{-g}\, dx^1 dx^2 dx^3 . \tag{28}$$

From comparison of (27) and (28) it follows that the three-dimensional quantity $\mathfrak{J}$ and the generalized momentum coincide: $\mathfrak{J} = \mathbf{p}$. The same equality was found at the end of the previous section using Lagrangian in (22). Thus, the Lagrangian density (23) and its part (24) allow us to calculate the generalized momentum of particles $\mathbf{p}$, which coincides with the relativistic momentum of the particles.

From (26), at $\mu = 0$, we find the time component of the generalized four-momentum:

$$p_0 = \frac{1}{c^2} \int\limits_{V_s} \left( \rho_{0q} \varphi + \rho_0 \psi + \rho_0 \vartheta + \rho_0 \wp \right) u^0 \sqrt{-g}\, dx^1 dx^2 dx^3 \ . \qquad (29)$$

It can be seen from the above that the four-vectors $\mathcal{P}_\mu$ and $p_\mu$ characterize the volumetric density and the total generalized four-momentum of all the particles, respectively, that is, they are calculated over the entire system's material. The contribution to these four-vectors is made by all the fields acting in the system. However, the fields are present not only in the material, but some of them also act outside of the material. Typical examples are electromagnetic and gravitational fields. If the system moves as a whole, then the fields outside the system acquire an additional four-momentum, which must be added to the generalized four-momentum $p_\mu$, if we want to find the total four-momentum of the system of particles and fields. Thus, the generalized four-momentum $p_\mu$ is only part of the total four-momentum of the system, while the time component $p_0$ defines the energy of particles and fields in the system's material, and the space component $p_i$ with the index $i = 1, 2, 3$ defines the relativistic (generalized) momentum of these particles and fields.

According to (18), in the equilibrium and closed system $p_\mu$ is conserved, and the same can be said about the four-momentum of electromagnetic and gravitational fields of the system outside the material, as well as about the total four-momentum of the system. The reason for conservation of the total four-momentum of a closed system is impossibility of the four-momentum's changing due to the lack of interaction with the environment, while it is assumed that the internal interactions are not able to change the system's four-momentum. The condition of equilibrium system implies that the proportions of energy and momentum for the particles and fields remain unchanged all the time, which ensures conservation of the generalized four-momentum $p_\mu$, as well as of the four-momentum of fields outside the material.

## 6. Relativistic uniform system at rest

We will consider within the framework of special theory of relativity (STR) a relativistic uniform system, which is closely filled with a multitude of particles and is held in equilibrium by four vector fields. For macroscopic bodies, the main acting force is the gravitational force, which gives the bodies a spherical shape.

Let us suppose that all the system's particles move randomly and independently of each other, and there are no directed fluxes of material and general rotation in the system. We will also assume that in the particles' comoving reference frames both the proper vector field potentials and the particles' solenoidal vectors vanish. Then, in the rest system, the potentials and field tensors will not depend on the four-velocities of individual particles and formulas (26-29) will be applicable.

For electromagnetic field, for example, this means that charged particles do not have their intrinsic magnetic moment in their comoving reference frames. As for acceleration field, the particles must have proper rotation close to zero. Under such assumptions, it is easy to show that as a result of solving wave equations for individual particles and for a great number of randomly moving particles in the system under consideration, the global vector potentials $\mathbf{A}$, $\mathbf{D}$, $\mathbf{U}$, $\mathbf{\Pi}$, as well as the solenoidal field vectors in the system tend to zero. This leads to the fact that $\mathbf{p}$ in (27) and $\mathfrak{I}$ in (28) become equal to zero, and it suffices for us to determine only the time component $p_0$ in (29). Within the framework of STR, in (29) $\sqrt{-g}=1$, the sphere's volume element $dx^1dx^2dx^3=dV_s$ and taking into account the time component of $\mathcal{P}_\mu$ in (25) we can write:

$$\mathcal{P}_0=\rho_{0q}A_0+\rho_0 D_0+\rho_0 U_0+\rho_0\pi_0=\frac{1}{c}\left(\rho_{0q}\varphi+\rho_0\psi+\rho_0\vartheta+\rho_0\wp\right). \qquad (30)$$

$$p_0=\frac{1}{c}\int_{V_s}\mathcal{P}_0 u^0\, dV_s\,. \qquad (31)$$

The scalar potentials of fields inside a sphere in the case $\rho_{0q}=const$ and $\rho_0=const$ were determined in [15-17]:

$$\varphi=\frac{c^2\rho_{0q}\gamma_c}{4\pi\varepsilon_0\eta\rho_0 r}\left[\frac{c}{\sqrt{4\pi\eta\rho_0}}\sin\left(\frac{r}{c}\sqrt{4\pi\eta\rho_0}\right)-r\cos\left(\frac{a}{c}\sqrt{4\pi\eta\rho_0}\right)\right]\approx\frac{\rho_{0q}\gamma_c(3a^2-r^2)}{6\varepsilon_0}.$$

$$\psi=-\frac{Gc^2\gamma_c}{\eta r}\left[\frac{c}{\sqrt{4\pi\eta\rho_0}}\sin\left(\frac{r}{c}\sqrt{4\pi\eta\rho_0}\right)-r\cos\left(\frac{a}{c}\sqrt{4\pi\eta\rho_0}\right)\right]\approx-\frac{2\pi G\rho_0\gamma_c(3a^2-r^2)}{3}.$$

$$\vartheta=\gamma' c^2, \qquad \gamma'=\frac{c\gamma_c}{r\sqrt{4\pi\eta\rho_0}}\sin\left(\frac{r}{c}\sqrt{4\pi\eta\rho_0}\right)\approx\gamma_c-\frac{2\pi\eta\rho_0 r^2\gamma_c}{3c^2}.$$

$$\wp = \wp_c - \frac{\sigma c^2 \gamma_c}{\eta} + \frac{\sigma c^3 \gamma_c}{r\eta\sqrt{4\pi\eta\rho_0}} \sin\left(\frac{r}{c}\sqrt{4\pi\eta\rho_0}\right) \approx \wp_c - \frac{2\pi\sigma\rho_0 r^2 \gamma_c}{3}. \qquad (32)$$

In (32) $\varepsilon_0$ is the electric constant, $\gamma_c$ is the Lorentz factor of particles at the center of sphere, $a$ is the sphere's radius, $\wp_c$ is the scalar potential of pressure field at the center of the sphere. For the charge four-current we have: $j^\mu = \rho_{0q} u^\mu$, while the four-velocity of the particles $u^\mu = (\gamma' c, \gamma' \mathbf{v}')$, $u^0 = \gamma' c$, where $\gamma' = \frac{1}{\sqrt{1 - v'^2/c^2}}$ is the Lorentz factor for the particles, $\mathbf{v}'$ is the root-mean-square velocity of the particles.

The appearance of sines and cosines in (32) is associated with taking into account the Lorentz factor of the proper chaotic motion of particles. If we neglect the internal motion of the particles, then the field potentials will become equal to the potentials inside an ideal solid sphere. Such potentials are indicated as approximate expressions on the right-hand sides in (32).

In [12], when analyzing equation of motion, it was shown that in the system under consideration the following relation between the field coefficients held true:

$$\eta + \sigma = G - \frac{\rho_{0q}^2}{4\pi\varepsilon_0 \rho_0^2}. \qquad (33)$$

Let us substitute potentials (32) into (30) and take into account (33):

$$\mathcal{P}_0 = \frac{\rho_0 c \gamma_c}{\eta} (\eta + \sigma) \cos\left(\frac{a}{c}\sqrt{4\pi\eta\rho_0}\right) + \frac{\rho_0}{c}\left(\wp_c - \frac{\sigma c^2 \gamma_c}{\eta}\right). \qquad (34)$$

We can write the density of generalized four-momentum in terms of components as follows: $\mathcal{P}_\mu = (\mathcal{P}_0, -\boldsymbol{\mathcal{P}})$, where $\boldsymbol{\mathcal{P}}$ is the density of three-dimensional generalized momentum and space component of the four-vector. In the case under consideration, it turns out that $\boldsymbol{\mathcal{P}} = 0$, and according to (34) $\mathcal{P}_0 = const$ in the entire volume of a sphere. Thus, $\mathcal{P}_\mu$ turns out to be a constant four-vector.

We will substitute $\mathcal{P}_0$ from (34) into (31) and will integrate over the sphere's volume. Since $u^0 = \gamma' c$, where $\gamma'$ is specified in (32), we obtain the following:

$p_0 =$

$$= \frac{ac\gamma_c}{\eta}\left[\frac{c^2\gamma_c}{\eta}(\eta+\sigma)\cos\left(\frac{a}{c}\sqrt{4\pi\eta\rho_0}\right)+\wp_c-\frac{\sigma c^2\gamma_c}{\eta}\right]\left[\frac{c\sin\left(\frac{a}{c}\sqrt{4\pi\eta\rho_0}\right)}{a\sqrt{4\pi\eta\rho_0}}-\cos\left(\frac{a}{c}\sqrt{4\pi\eta\rho_0}\right)\right].$$

(35)

If in (35) the inequality $\frac{a}{c}\sqrt{4\pi\eta\rho_0}\ll 1$ holds true, then the sines and cosines can be expanded up to the second-order terms. This gives the following:

$$p_0 \approx \frac{4\pi\rho_0 a^3\gamma_c}{3c}\left(c^2\gamma_c+\wp_c\right)\approx\frac{m\gamma_c}{c}\left(c^2\gamma_c+\wp_c\right).$$

where $m=\frac{4\pi\rho_0 a^3}{3}$ is a quantity with the dimension of mass, which is equal to the product of mass density $\rho_0$ by the sphere's volume.

On the other hand, it was shown in [15] that the total mass of particles inside a sphere is defined by the quantity $m_b$, which differs from $m$. The difference in masses arises from the particles' motion, since the effective density of a moving particle equals $\rho_0\gamma'$. The total mass of particles inside the sphere is defined by the integral over the sphere's volume:

$$m_b = m_g = \int dm = \int \rho_0\gamma' dV = \frac{c^2\gamma_c}{\eta}\left[\frac{c}{\sqrt{4\pi\eta\rho_0}}\sin\left(\frac{a}{c}\sqrt{4\pi\eta\rho_0}\right)-a\cos\left(\frac{a}{c}\sqrt{4\pi\eta\rho_0}\right)\right].$$

(36)

Furthermore, it turns out that the mass $m_b$ is equal to the gravitational mass $m_g$, which specifies the scalar potential and the gravitational field strength outside the sphere. For the charge $q_b$ of the system, similarly to (36), we find:

$$q_b = \int dq = \int \rho_{0q}\gamma' dV = \frac{\rho_{0q}c^2\gamma_c}{\eta\rho_0}\left[\frac{c}{\sqrt{4\pi\eta\rho_0}}\sin\left(\frac{a}{c}\sqrt{4\pi\eta\rho_0}\right)-a\cos\left(\frac{a}{c}\sqrt{4\pi\eta\rho_0}\right)\right].$$

Let us substitute (36) into (35):

$$p_0 = \frac{m_b}{c}\left[\frac{c^2\gamma_c}{\eta}(\eta+\sigma)\cos\left(\frac{a}{c}\sqrt{4\pi\eta\rho_0}\right)+\wp_c-\frac{\sigma c^2\gamma_c}{\eta}\right]\approx\frac{m_b}{c}\left(c^2\gamma_c+\wp_c\right). \qquad (37)$$

Hence, we can see that the time component $p_0$ of the system's generalized four-momentum exceeds the value $m_b c\gamma_c$ by approximately $\frac{m_b}{c}\wp_c$. We will write the four-vector $p_\mu$ in terms of time and space components: $p_\mu = (p_0, -\mathbf{p})$, where $\mathbf{p}$ is a three-dimensional generalized momentum. The four-vector $p_\mu$ can be considered a constant four-vector, since according to (37) $p_0 = const$ as long as the Lorentz factor $\gamma_c$ and the scalar potential $\wp_c$ at the center of sphere are constant, which is true for an equilibrium system. In addition, according to (16) and (26) $\mathbf{p}=0$ as a consequence of the fact that $\boldsymbol{\mathcal{P}}=0$ in the definition $\mathcal{P}_\mu = (\mathcal{P}_0, -\boldsymbol{\mathcal{P}})$.

Since the four-vector $\mathcal{P}_\mu$ and $p_\mu$ for the sphere at rest turn out to be constant, then relation (18) holds true, and (13) implies the following:

$$\frac{\partial}{\partial x^\mu}\left(\frac{\mathcal{L}_p}{u^0}\right)=0. \qquad (38)$$

Let us verify relation (38) for the case of the sphere at rest within the framework of STR. For this, it is necessary to express the relation $\mathcal{L}_p / u^0$ in terms of components of the four-radius $x^\mu = (ct, x, y, z)$. We will consider the sum of products of the fields' four-potentials by the four-currents in (24), and will express this sum in terms of its components. Thus, for the electromagnetic field and other fields inside the sphere we will obtain the following:

$$A_\mu = \left(\frac{\varphi}{c}, -\mathbf{A}\right), \qquad j^\mu = \rho_{0q} u^\mu = \rho_{0q}(\gamma' c, \gamma' \mathbf{v}), \qquad A_\mu\, j^\mu = \gamma'\rho_{0q}\varphi - \gamma'\rho_{0q}\mathbf{A}\cdot\mathbf{v}.$$

$$D_\mu = \left(\frac{\psi}{c}, -\mathbf{D}\right), \qquad J^\mu = \rho_0 u^\mu = \rho_0(\gamma' c, \gamma' \mathbf{v}), \qquad D_\mu\, J^\mu = \gamma'\rho_0\psi - \gamma'\rho_0\mathbf{D}\cdot\mathbf{v}.$$

$$U_\mu\, J^\mu = \gamma'\rho_0\vartheta - \gamma'\rho_0\mathbf{U}\cdot\mathbf{v}, \qquad \pi_\mu\, J^\mu = \gamma'\rho_0\wp - \gamma'\rho_0\mathbf{\Pi}\cdot\mathbf{v}. \qquad (39)$$

Since the fields' vector potentials $\mathbf{A}$, $\mathbf{D}$, $\mathbf{U}$, $\mathbf{\Pi}$ in this case are equal to zero, we can write:

$$A_\mu \, j^\mu + D_\mu \, J^\mu + U_\mu \, J^\mu + \pi_\mu \, J^\mu = \gamma' \rho_{0q} \varphi + \gamma' \rho_0 \psi + \gamma' \rho_0 \vartheta + \gamma' \rho_0 \wp .$$

Let us substitute here (30) and take into account that $u^0 = \gamma' c$:

$$A_\mu \, j^\mu + D_\mu \, J^\mu + U_\mu \, J^\mu + \pi_\mu \, J^\mu = u^0 \mathcal{P}_0 = -\mathcal{L}_p . \qquad (40)$$

According to (34) $\mathcal{P}_0 = const$, so in view of (40), relation (38) holds true:

$$\frac{\partial}{\partial x^\mu} \left( \frac{\mathcal{L}_p}{u^0} \right) = -\frac{\partial \mathcal{P}_0}{\partial x^\mu} = 0 .$$

## 7. Moving relativistic uniform system

Let us consider a sphere with the particles moving at a constant velocity $\mathbf{V}$ along the axis $OX$, and at initial time point the center of the sphere was located at the origin of fixed reference frame $K$. In the reference frame $K'$, associated with the center of the sphere, the scalar potentials of the fields are expressed by formulas (32), and the vector potentials of the fields on the average are equal to zero.

We can determine the field potentials from the standpoint of the reference frame $K$, taking into account the fact that field potentials are part of the corresponding four-potentials, which are transformed from $K'$ into $K$ as four-vectors. Within the framework of SRT, the four-potentials are transformed in the same way as the time and coordinates in the Lorentz transformations. For example, if the four-potential of electromagnetic field in $K'$ is $A'_\mu = \left( \frac{\varphi'}{c}, -A'_x, -A'_y, -A'_z \right)$, then in $K$ for the components of the four-potential we can write the following:

$$A_\mu = \left( \frac{\varphi}{c}, -A_x, -A_y, -A_z \right) = \left[ \frac{\gamma}{c} \left( \varphi' + V \cdot A'_x \right), -\gamma \left( A'_x + V \varphi' / c^2 \right), -A'_y, -A'_z \right] =$$
$$= \left( \frac{\gamma \varphi'}{c}, -\frac{\gamma \varphi' V}{c^2}, 0, 0 \right). \qquad (41)$$

Here $\gamma=\frac{1}{\sqrt{1-V^2/c^2}}$ is the Lorentz factor of motion of the sphere's center in $K$.

In (41) it is taken into account that in $K'$, where the sphere is motionless, all the three components of the vector potential $\mathbf{A}'=(A_1', A_2', A_3')$ are equal to zero. For the four-potentials of other fields we can write in a similar way:

$$D_\mu=\left(\frac{\psi}{c},-D_x,-D_y,-D_z\right)=\left(\frac{\gamma\psi'}{c},-\frac{\gamma\psi' V}{c^2},0,0\right).$$

$$U_\mu=\left(\frac{\vartheta}{c},-U_x,-U_y,-U_z\right)=\left(\frac{\gamma\vartheta'}{c},-\frac{\gamma\vartheta' V}{c^2},0,0\right).$$

$$\pi_\mu=\left(\frac{\wp}{c},-\Pi_x,-\Pi_y,-\Pi_z\right)=\left(\frac{\gamma\wp'}{c},-\frac{\gamma\wp' V}{c^2},0,0\right). \qquad (42)$$

In $K'$ the velocity of an arbitrary particle inside the sphere equals $\mathbf{v}'$, and the Lorentz factor is $\gamma'=\frac{1}{\sqrt{1-v'^2/c^2}}$. Let us denote the total velocity of the particle in $K$ by $\mathbf{v}$ and the Lorentz factor of the particle by $\gamma_p=\frac{1}{\sqrt{1-v^2/c^2}}$. Transforming the particle's four-velocity from $K'$ into $K$ using the Lorentz transformations gives the following:

$$u^\mu=\left(\gamma_p c,\gamma_p v_x,\gamma_p v_y,\gamma_p v_z\right)=\left[c\gamma\gamma'\left(1+Vv_x'/c^2\right),\gamma\gamma'\left(v_x'+V\right),\gamma' v_y',\gamma' v_z'\right]. \qquad (43)$$

Let us substitute (41) and (42) into (25) and find in $K$ the time and space components of the density of generalized four-momentum $\mathcal{P}_\mu=(\mathcal{P}_0,-\boldsymbol{\mathcal{P}})$, where $\boldsymbol{\mathcal{P}}=(\mathcal{P}_x,\mathcal{P}_y,\mathcal{P}_z)$:

$$\mathcal{P}_0=\rho_{0q}A_0+\rho_0 D_0+\rho_0 U_0+\rho_0\pi_0=\frac{\gamma}{c}\left(\rho_{0q}\varphi'+\rho_0\psi'+\rho_0\vartheta'+\rho_0\wp'\right).$$

$$\mathcal{P}_x=\rho_{0q}A_x+\rho_0 D_x+\rho_0 U_x+\rho_0\Pi_x=\frac{\gamma V}{c^2}\left(\rho_{0q}\varphi'+\rho_0\psi'+\rho_0\vartheta'+\rho_0\wp'\right)=\frac{V}{c}\mathcal{P}_0.$$

$$\mathcal{P}_y = \mathcal{P}_z = 0 \,. \qquad (44)$$

In (44), the fields' scalar potentials $\varphi'$, $\psi'$, $\vartheta'$ and $\wp'$ in the reference frame $K'$ are the scalar potentials, which are presented in (32). With this in mind, we can use expressions (30), (33) and (34) for $K'$, and for the reference frame $K$ we find:

$$\mathcal{P}_0 = \frac{\rho_0 \, c\gamma_c \gamma}{\eta} (\eta + \sigma) \cos\left( \frac{a}{c} \sqrt{4\pi \eta \rho_0} \right) + \frac{\rho_0 \gamma}{c} \left( \wp_c - \frac{\sigma c^2 \gamma_c}{\eta} \right).$$

$$\mathcal{P}_x = \frac{V}{c} \mathcal{P}_0 \,, \qquad\qquad \mathcal{P}_y = \mathcal{P}_z = 0 \,. \qquad (45)$$

According to (45), in $K$ the time component of the density of generalized four-momentum increases by a factor of $\gamma$ as compared to $K'$. In addition, the component $\mathcal{P}_x$ of the density of three-dimensional generalized momentum along the axis $OX$ appears, which is proportional to the velocity of the sphere's motion in $K$.

In order to simplify the calculations, we will assume that the sphere's velocity $V$ significantly exceeds the particles' velocities $v'_x$, $v'_y$ and $v'_z$, so the latter can be neglected. If in (43) $Vv'_x \ll c^2$, then the time component $u^0 \approx c\gamma\gamma'$. Taking from (45) the components $\mathcal{P}_\mu = (\mathcal{P}_0, -\boldsymbol{\mathcal{P}})$ in $K$, with the help of (16), (26-27) at $\sqrt{-g} = 1$ we can determine the components of the generalized four-momentum $p_\mu = (p_0, -\mathbf{p})$, where $\mathbf{p} = (p_x, p_y, p_z)$:

$$p_\mu = \frac{1}{c} \int_{V_s} \mathcal{P}_\mu u^0 \, dx^1 dx^2 dx^3 \,, \qquad p_0 = \frac{1}{c} \int_{V_s} \mathcal{P}_0 u^0 \, dx^1 dx^2 dx^3 \approx \mathcal{P}_0 \gamma \int_{V_s} \gamma' \, dx^1 dx^2 dx^3 \,.$$

$$\mathbf{p} = \frac{1}{c} \int_{V_s} \boldsymbol{\mathcal{P}} u^0 \, dx^1 dx^2 dx^3 \approx \gamma \int_{V_s} \boldsymbol{\mathcal{P}} \gamma' \, dx^1 dx^2 dx^3 \,.$$

$$p_x \approx \gamma \int_{V_s} \mathcal{P}_x \gamma' \, dx^1 dx^2 dx^3 = \frac{\mathcal{P}_0 \gamma V}{c} \int_{V_s} \gamma' \, dx^1 dx^2 dx^3 = \frac{p_0 V}{c} \,, \qquad p_y = p_z = 0 \,. \qquad (46)$$

If in (46) we calculate the time component $p_0$ of generalized four-momentum, then the component $p_x$ of three-dimensional generalized momentum would be thereby determined.

According to STR approach, a moving sphere with the particles is represented in $K$ as a Heaviside ellipsoid, regardless of the internal motion of particles in $K'$. In [18], the energy and momentum of electromagnetic field of a moving charged sphere were studied and the 4/3 problem was discovered. The same was discovered in [19] for the gravitational field. Next, we will proceed similarly to [18-19], and will introduce in $K$ new coordinates $r,\theta,\varphi$ associated with the Cartesian coordinates:

$$x-Vt=\frac{1}{\gamma}r\cos\theta\,,\qquad y=r\sin\theta\cos\varphi\,,\qquad z=r\sin\theta\sin\varphi\,. \tag{47}$$

In these coordinates, the volume element in (46) is determined by the formula $dx^1dx^2dx^3=\frac{1}{\gamma}r^2\sin\theta\,dr\,d\theta\,d\varphi$. According to (32), in the reference frame $K'$ the Lorentz factor $\gamma'$ of particles moving inside the sphere is expressed in terms of a current radius, which we will denote here by $r'$:

$$\gamma'=\frac{c\gamma_c}{r'\sqrt{4\pi\eta\rho_0}}\sin\left(\frac{r'}{c}\sqrt{4\pi\eta\rho_0}\right).$$

If we take into account the Lorentz transformations, then the coordinates $r,\theta,\varphi$ in $K$ inside of the Heaviside ellipsoid present in (47) coincide with the spherical coordinates $r',\theta',\varphi'$ in $K'$ inside of the sphere, so $r'=\sqrt{\gamma^2\left(x-vt\right)^2+y^2+z^2}=r$.

All this allows us to calculate the integral for $p_0$ in (46):

$$p_0\approx\mathcal{P}_0\gamma\int_{V_s}\gamma'\,dx^1dx^2dx^3=\frac{c\gamma_c}{\sqrt{4\pi\eta\rho_0}}\mathcal{P}_0\int_{V_s}\sin\left(\frac{r}{c}\sqrt{4\pi\eta\rho_0}\right)r\sin\theta\,dr\,d\theta\,d\varphi=$$
$$=\frac{c^2\gamma_c}{\eta\rho_0}\mathcal{P}_0\left[\frac{c}{\sqrt{4\pi\eta\rho_0}}\sin\left(\frac{a}{c}\sqrt{4\pi\eta\rho_0}\right)-a\cos\left(\frac{a}{c}\sqrt{4\pi\eta\rho_0}\right)\right].$$

Let us substitute here $\mathcal{P}_0$ from (45) and $m_b$ from (36):

$$p_0 \approx \frac{\gamma m_b}{c}\left[\frac{c^2\gamma_c}{\eta}(\eta+\sigma)\cos\left(\frac{a}{c}\sqrt{4\pi\eta\rho_0}\right)+\wp_c-\frac{\sigma c^2\gamma_c}{\eta}\right]\approx\frac{\gamma m_b}{c}\left(c^2\gamma_c+\wp_c\right). \qquad (48)$$

In comparison with (37), the component $p_0$ has increased by a factor of $\gamma$ due to the motion of the physical system as a whole at velocity $V$.

For the component $p_x$ of the generalized momentum from (46) we find:

$$p_x \approx \frac{p_0 V}{c}, \qquad (49)$$

where the component $p_0$ is calculated in (48).

As long as the sphere with the particles moves at a constant velocity $V$ along the axis $OX$, we can assume that in (48) is the component $p_0 = const$. The same will also be true for $p_x$ in (49), while $p_y = p_z = 0$. Hence it follows that the generalized four-momentum $p_\mu = (p_0, -\mathbf{p})$, where $\mathbf{p} = (p_x, p_y, p_z)$, is a constant four-vector, and therefore the conservation condition (18) holds true.

According to definition of the four-potential of pressure field in [6], for scalar potential at the center of a sphere we can write: $\wp_c = \frac{\gamma_c\, p_{oc}}{\rho_0}$, where $p_{oc}$ is the proper pressure inside a typical particle moving at the center of the sphere. With this in mind, the expression for momentum of the system's particles follows from (48) and (49):

$$\mathbf{p} \approx \gamma\gamma_c m_b\left(1+\frac{p_{oc}}{\rho_0 c^2}\right)\mathbf{V}.$$

As we can see, taking into account the proper pressure $p_{oc}$ and the proper density $\rho_0$ of particles increases the value of the total momentum of the system's particles, regardless of the contribution of the Lorentz factors $\gamma$ and $\gamma_c$ to the momentum.

According to (48) and (49), the generalized four-momentum can be written as follows:

$$p_\mu = p_0\left(1, -\frac{\mathbf{V}}{c}\right) \approx \frac{m_b}{c^2}\left(c^2\gamma_c+\wp_c\right)\left(\gamma c, -\gamma\mathbf{V}\right) \approx m_b\left(\gamma_c+\frac{\wp_c}{c^2}\right)u_\mu, \qquad (50)$$

where $u_\mu$ is four-velocity of the center of a sphere. Thus, the generalized 4-momentum is directed along the four-velocity of the system under consideration.

From (48-50) it follows

$$\mathbf{p} \approx \gamma m_b \left( \gamma_c + \frac{\wp_c}{c^2} \right) \mathbf{V}$$

so that in the first approximation the total momentum of particles is proportional to the Lorentz factor $\gamma$, the velocity $\mathbf{V}$ of the center of momentum's motion and the total mass $m_b$ of the system's particles defined in (36). Besides, the greater are the scalar potential $\vartheta_c = c^2 \gamma_c$ of acceleration field and the scalar potential $\wp_c$ of pressure field at the center of a sphere, the greater is the momentum. Since $\gamma_c$ is the Lorentz factor of particles at the center of the sphere, we can see that due to the motion of particles inside the sphere, the effective mass, which is included in the momentum of particles of the system, increases. This means that instead of the mass $m_b$, which is typical for a resting relativistic uniform system, the value $m_b \left( \gamma_c + \frac{\wp_c}{c^2} \right)$ becomes the effective total mass of particles in the moving system.

It remains for us to verify Equation (13). From (45) it follows that $\mathcal{P}_0 = const$, $\mathcal{P}_x = \frac{V}{c} \mathcal{P}_0 = const$, $\mathcal{P}_y = \mathcal{P}_z = 0$. This means that the density of generalized four-momentum $\mathcal{P}_\mu = (\mathcal{P}_0, -\boldsymbol{\mathcal{P}})$, where $\boldsymbol{\mathcal{P}} = (\mathcal{P}_x, \mathcal{P}_y, \mathcal{P}_z)$, is a constant four-vector, and then the left-hand side of Equation (13) becomes equal to zero, $\frac{d\mathcal{P}_\mu}{dt} = 0$. We will consider the right-hand side of (13), which contains the value $-c \frac{\partial}{\partial x^\mu} \left( \frac{\mathcal{L}_p}{u^0} \right)$. Using the expression for $\mathcal{L}_p$ in (24), we find:

$$\begin{aligned} \frac{\mathcal{L}_p}{u^0} &= \frac{1}{u^0} \left( -A_\mu \, j^\mu - D_\mu \, J^\mu - U_\mu \, J^\mu - \pi_\mu \, J^\mu \right) = \\ &= -\frac{1}{c} \left( \rho_{0q} \varphi - \rho_{0q} \mathbf{A} \cdot \mathbf{v} + \rho_0 \psi - \rho_0 \mathbf{D} \cdot \mathbf{v} + \rho_0 \vartheta - \rho_0 \mathbf{U} \cdot \mathbf{v} + \rho_0 \wp - \rho_0 \mathbf{\Pi} \cdot \mathbf{v} \right). \end{aligned} \qquad (51)$$

According to (43), the Lorentz factor $\gamma_p$ and the components of total velocity $\mathbf{v}=(v_x,v_y,v_z)$ of an arbitrary particle during motion of a sphere with the particles in $K$ equal:

$$\gamma_p=\gamma\gamma'\left(1+Vv'_x/c^2\right), \qquad v_x=\frac{\gamma\gamma'\left(v'_x+V\right)}{\gamma_p}=\frac{v'_x+V}{1+Vv'_x/c^2},$$

$$v_y=\frac{\gamma' v'_y}{\gamma_p}=\frac{v'_y}{\gamma\left(1+Vv'_x/c^2\right)}, \qquad v_z=\frac{\gamma' v'_z}{\gamma_p}=\frac{v'_z}{\gamma\left(1+Vv'_x/c^2\right)}. \tag{52}$$

If in (52) we neglect the components $v'_x$, $v'_y$, $v'_z$ of the particle's proper velocity inside the sphere measured in $K'$, then it will be $\mathbf{v}\approx(V,0,0)$. Then for the electromagnetic field $\mathbf{A}\cdot\mathbf{v}\approx A_x V=\frac{\gamma\varphi' V^2}{c^2}$, and similar expressions will hold for the other fields, in view of (41) and (42). We will substitute this into (51) and will take into account the expressions for scalar potentials of the form $\varphi=\gamma\varphi'$ from (41) and (42), as well as $\mathcal{P}_0$ from (44):

$$\frac{\mathcal{L}_p}{u^0}=-\frac{\gamma}{c}\left(1-\frac{V^2}{c^2}\right)\left(\rho_{0q}\varphi'+\rho_0\psi'+\rho_0\vartheta'+\rho_0\wp'\right)=$$
$$=-\frac{1}{\gamma c}\left(\rho_{0q}\varphi'+\rho_0\psi'+\rho_0\vartheta'+\rho_0\wp'\right)=-\frac{\mathcal{P}_0}{\gamma^2}.$$

Since according to (45) $\mathcal{P}_0=const$, we obtain the value $\frac{\mathcal{L}_p}{u^0}=const$. Consequently, the right-hand side of (13) will be equal to zero, that is, $-c\frac{\partial}{\partial x^\mu}\left(\frac{\mathcal{L}_p}{u^0}\right)=0$, and Equation (13) is satisfied.

### 8. Discussion

When we calculated the generalized four-momentum of a moving uniform relativistic system in the reference frame $K$, instead of the time component of four-velocity of an arbitrary particle $u^0=c\gamma\gamma'\left(1+Vv'_x/c^2\right)$ in (43) we used an approximate value $u^0\approx c\gamma\gamma'$. This led to the fact that the time component $p_0$ in (48) increased by a factor of $\gamma$ due to the motion of the physical system as a whole at the velocity $V$ as compared to the static case. Will anything change if we take into account the velocity component $v'_x$ in the expression for $u^0$? In an equilibrium system of particles, which is stationary in general, the total momentum of these particles as a rule is equal to zero. When the particles move randomly, their momenta are subtracted from each other due to the different directions of the particles' velocities, the same is true for freely rotating systems. In addition, in the center-of-momentum frame the total momentum is always equal to zero. The velocity component $v'_x$ is included in $u^0$ as an additive raised to the first odd power, and then is integrated over the volume when we calculate $p_\mu$ in (46). This additive behaves as a certain antisymmetric function changing its sign, the volume integral of which becomes equal to zero. Therefore, the estimates of $p_0$ and $p_x$ obtained in (48) and (49) remain unchanged.

The relativistic energy $E$ for a system of particles and vector fields was found in [5] in a curved space-time. If the system is stationary and there is no energy dissipation due to non-potential forces, then the Hamiltonian $H$ of the system becomes equal to the energy:

$$H = E = \frac{1}{c} \int_{V_s} \left( \rho_0 \psi + \rho_{0q} \varphi + \rho_0 \vartheta + \rho_0 \wp \right) u^0 \sqrt{-g}\, dx^1 dx^2 dx^3 - \\ - \int_{V_s} \begin{pmatrix} ckR - 2ck\Lambda + \frac{c^2}{16\pi G} \Phi_{\mu\nu} \Phi^{\mu\nu} - \frac{1}{4\mu_0} F_{\mu\nu} F^{\mu\nu} - \\ - \frac{c^2}{16\pi\eta} u_{\mu\nu} u^{\mu\nu} - \frac{c^2}{16\pi\sigma} f_{\mu\nu} f^{\mu\nu} \end{pmatrix} \sqrt{-g}\, dx^1 dx^2 dx^3. \qquad (53)$$

We substitute $\mathcal{L}$ from (23) into (1), using the energy calibration condition in the form $ckR - 2ck\Lambda = 0$ according to [5], [20], add the result for $L$ with $H$ (53) and take into account (25):

$$L + H = -\int_{V_s} \mathcal{P}_\mu u^\mu \sqrt{-g}\, dx^1 dx^2 dx^3 + \int_{V_s} \mathcal{P}_0\, u^0 \sqrt{-g}\, dx^1 dx^2 dx^3 = -\int_{V_s} \mathcal{P}_i u^i \sqrt{-g}\, dx^1 dx^2 dx^3 .$$

Here the index $i = 1, 2, 3$ defines spatial components of four-vectors $\mathcal{P}_\mu$ and $u^\mu$. We now take into account that $\mathcal{P}_\mu = (\mathcal{P}_0, -\boldsymbol{\mathcal{P}})$, and four-velocity $u^\mu = \frac{dx^\mu}{d\tau} = \frac{1}{c} u^0 \frac{dx^\mu}{dt} = \frac{1}{c} u^0 (c, \mathbf{v})$:

$$L+H=\frac{1}{c}\int\limits_{V_s}\mathcal{P}\cdot\mathbf{v}u^0\sqrt{-g}\,dx^1dx^2dx^3 .$$

Using (17), we can replace volumes of moving particles with their proper volumes and replace the integral over volume with the sum of integrals over volumes of individual particles:

$$L+H=\int\limits_{V_{s0}}\mathcal{P}\cdot\mathbf{v}\,dV_0=\sum_{n=1}^{N}\int\limits_{V_{0n}}\mathcal{P}_n\cdot\mathbf{v}_n\,dV_{0n}=\sum_{n=1}^{N}\mathbf{v}_n\cdot\int\limits_{V_{0n}}\mathcal{P}_n\,dV_{0n} .$$

Since $p_\mu=(p_0,-\mathbf{p})$, in view of (16) we find:

$$L+H=\sum_{n=1}^{N}\mathbf{v}_n\cdot\mathbf{p}_n ,$$

This expression is a standard Legendre transformation connecting the Lagrangian, Hamiltonian, velocities and generalized three-dimensional momenta of all particles of the system. Thus, the concept of the generalized four-momentum presented by us is consistent both with Hamiltonian mechanics and Lagrange mechanics [21].

From (53), on condition of energy gauging in the form $ckR-2ck\Lambda=0$, and from (29) it follows:

$$E=cp_0-\int\limits_{V_s}\left(\begin{array}{l}\frac{c^2}{16\pi G}\Phi_{\mu\nu}\Phi^{\mu\nu}-\frac{1}{4\mu_0}F_{\mu\nu}F^{\mu\nu}-\\-\frac{c^2}{16\pi\eta}u_{\mu\nu}u^{\mu\nu}-\frac{c^2}{16\pi\sigma}f_{\mu\nu}f^{\mu\nu}\end{array}\right)\sqrt{-g}\,dx^1dx^2dx^3 . \qquad (54)$$

This means that the time component $p_0$ of the generalized four-momentum of a system defines a part of the energy-momentum that is associated with the particles affected by the system's fields. As for contribution of the fields themselves to the system's energy, it is defined by the integral in (54), according to [15], [16], [22]. We can assume that separation of energy in (54) into particle energy and field energy arises from the very structure of the Lagrangian density (23). In this Lagrangian density, there is part (24) containing four-potentials of fields and four-currents of particles, and there is also a part containing tensor invariants appearing in the integral in (54).

Let us also consider the approach to the problem in question within the framework of the general theory of relativity (GTR). According to [13], [21], the Lagrangian density of GTR for the relativistic fluid can be represented as follows:

$$\mathcal{L}_{GTR}=-c\rho_0\sqrt{g_{\mu\nu}u^\mu u^\nu}-\frac{\rho_0}{c}\Pi\sqrt{g_{\mu\nu}u^\mu u^\nu}-\rho_{0q}A_\mu u^\mu-\frac{1}{4\mu_0}F_{\mu\nu}F^{\mu\nu}+ckR-2ck\Lambda\,. \qquad (55)$$

The function $\Pi=\int_0^p\frac{dp}{\rho_0}-\frac{p}{\rho_0}$ in (55) is the potential energy of elastic compression of the fluid per unit mass, and $p$ represents the pressure. The first three terms in (55) directly depend on the four-velocity $u^\mu$, and we can assume that they form that part of the Lagrangian $\mathcal{L}_p$, with the help of which the generalized momentum density $\mathcal{P}_\mu$ is calculated in (13). Hence we find:

$$\mathcal{P}_\mu=-\frac{\partial\mathcal{L}_p}{\partial u^\mu}=\rho_0 u_\mu+\frac{\rho_0}{c}\Pi u_\mu+\rho_{0q}A_\mu\,. \qquad (56)$$

Expression (56) for the generalized momentum density in GTR shows a significant difference in comparison with expression (25) obtained for the vector fields. In (56) the first term $\rho_0 u_\mu$ corresponds to the term $\rho_0 U_\mu$ in (25). However, the four-potential $U_\mu$ of the acceleration field is equal to the four-velocity $u_\mu$ only for a point particle, and in the general case for a fluid, as for a system of closely interacting particles, the inequality $U_\mu\neq u_\mu$ holds true [23]. The second term $\frac{\rho_0}{c}\Pi u_\mu$ in (56), associated with the pressure energy, corresponds to the term $\rho_0\pi_\mu$ in (25). But the term $\frac{\rho_0}{c}\Pi u_\mu$ is always directed along the four-velocity $u_\mu$, as for a free point particle, while actually the fluid particles interact with each other in such a way that the four-potential $\pi_\mu$ of the pressure field would always differ from the value $\frac{1}{c}\Pi u_\mu$. Finally, if the term $\rho_{0q}A_\mu$ for the electromagnetic field is identically represented in (25) and (56), then for the gravitational field the difference again is observed. In (25), the contribution to the generalized momentum density is made by $\rho_0 D_\mu$, where $D_\mu$ is the gravitational four-potential. But in (56) in the expression for $\mathcal{P}_\mu$ there isn't any term defining the gravitational field. This is an obvious consequence of the axiomatic of GTR, in which the spacetime metric plays the role of the gravitational field. Nevertheless, such equations as (12) and (13), into which the physical quantities averaged over typical particles should be substituted, must remain valid in GTR. This is possible, since the generalized four-force $\mathcal{F}_\mu$ in (13) depends on the metric and therefore on the gravitational field in GTR.

On the other hand, according to (16), the generalized four-momentum $p_\mu$ of system of particles is the volume integral of $\mathcal{P}_\mu$. Then it turns out that $p_\mu$ in GTR does not contain a contribution from the gravitational field, and therefore the space component $p_\mu$ cannot define the relativistic momentum of the particles of the system, in contrast to what we found for the vector fields in Section 4. The situation in GTR is made more complicated by the fact that an attempt to determine the four-momentum and the relativistic momentum of a physical system in another way, with the help of the volume integral of the time components of stress-energy tensor, even taking into account the gravitational field pseudotensor, is unsuccessful (see [21] and the references therein). Instead of the four-momentum, the so-called integral four-dimensional vector is obtained in this way, which characterizes distribution of energy and field energy fluxes in the system, is conserved in a closed system, but is not a standard locally defined four-vector.

## 9. Conclusion

The analysis of Lagrangian and its variation in the principle of least action has led us to the four-dimensional Euler-Lagrange Equation (12) and its variant (13) for the continuously distributed materials. In (16) we determine the generalized four-momentum $p_\mu = (p_0, -\mathbf{p})$, in (20) – an auxiliary four-dimensional quantity $\mathfrak{I}_\mu = (\mathfrak{I}_0, -\boldsymbol{\mathfrak{I}})$, in (21) the vector $\boldsymbol{\mathfrak{I}}$ and in (22) – the total relativistic momentum of particles of a system, found through the Lagrangian. By its definition, the generalized four-momentum $p_\mu$ turns out to be an integral four-vector, belonging to the special class of non-local four-vectors. As is shown at the end of Section 4, for such four-vectors a different order of transformation between the form with a covariant index and the form with a contravariant index is required.

Within the framework of the accepted assumptions, when for each particle $u^0$ does not change at moment the momentum is calculated, it turns out that $\mathbf{p} = \boldsymbol{\mathfrak{I}}$, moreover, $\mathbf{p}$ is equal to the total relativistic momentum of particles of the system.

Below, as an example, we use in (23) the Lagrangian density $\mathcal{L}$, which describes the relativistic vector fields, and in (24) its part $\mathcal{L}_p$, containing four-currents. As follows from definition of the generalized four-momentum, for its calculation it suffices to specify a part of the Lagrangian density $\mathcal{L}_p$. We calculate in terms of $\mathcal{L}_p$ the density of generalized four-momentum $\mathcal{P}_\mu = -\frac{\partial \mathcal{L}_p}{\partial u^\mu}$ in (25), as well as the terms of Equation (13). As a result, it turns out that for the vector fields the generalized four-momentum $p_\mu$ and the four-dimensional quantity $\Im_\mu$ coincide with each other, and Equation (13) is also satisfied in case of the system's motion at a constant velocity.

The results obtained are applied to uniform relativistic system in the form of a sphere, studied earlier in [24]. First, the components $p_0$ and $\mathbf{p}$, which are part of the generalized four-momentum $p_\mu$, are calculated for the system at rest, and then for the same system moving at a constant velocity. It follows from (37) and (48) that the component $p_0$ of the moving system is $\gamma$ times greater than the component $p_0$ of the resting system, where $\gamma$ is the Lorentz factor of motion of the center of sphere in the laboratory frame of reference. In this case, the moving system acquires a relativistic momentum $\mathbf{p}$ (49). A feature of the components $p_0$ and $\mathbf{p}$ is that in the first approximation they depend on the Lorentz factor $\gamma_c$ and on the potential $\wp_c$ of pressure field at the center of sphere. This can be seen in (50), where the generalized four-momentum is expressed in terms of four-velocity of the sphere.

Analysis of the current situation in general theory of relativity (GTR) shows that due to absence of covariant representation of contribution from the gravitational field, in relativistic hydrodynamics there is no complete description of relativistic and generalized four-momenta with the help of GTR. Available works are confined to the fact that the pressure has static nature, so neither the four-potential nor the pressure field tensor in the covariant formulation are used in description of the pressure field. The same is true for acceleration field, which not only defines the particles' energy density in the Lagrangian according to Einstein's formula, but also describes contribution from the energy of particles' own motion inside a system in terms of its four-potential and acceleration tensor. Instead, a phenomenological thermodynamic approach is usually used, in which the fluxes and the energies of particles are calculated in terms of temperature, pressure, entropy, chemical potential, etc. [25-34]. However, the approach based on the field theory and Lagrangian mechanics allows us to derive more convenient and covariant expressions for the generalized four-momentum $p_\mu$ in (16) and the generalized four-momentum density $\mathcal{P}_\mu$ in (25), which are valid in the curved spacetime. The results obtained are made possible by using the concept of typical particles to describe a continuous material, which makes it possible to simplify variation procedure and implement it completely in a four-dimensional form.